%% file: article.tex
\documentclass[12pt]{article}
\usepackage{epsfig}

\input econfmacros.tex

\RequirePackage{xspace}
\usepackage{relsize}

\usepackage{graphics}  
\usepackage{makeidx}
\usepackage{colordvi}
\usepackage{amsmath}
\usepackage{multirow}
\def\pep2{PEP-II \xspace}

\def\babar{\mbox{\slshape B\kern-0.1em{\smaller A}\kern-0.1em
    B\kern-0.1em{\smaller A\kern-0.2em R}}\xspace}

\def\invfb   {\ensuremath{\mbox{\,fb}^{-1}}\xspace}
\def\fs   {\ensuremath{\rm \,fs}\xspace}

\def\epem       {\ensuremath{e^+e^-}\xspace}

\def\CP                {\ensuremath{C\!P}\xspace}
\def\CPV                {\ensuremath{C\!PV}\xspace}

\def\FB                {{\scriptscriptstyle \rm FB}\xspace}

\def\Y#1S{\ensuremath{\Upsilon{(#1S)}}\xspace}

\def\Bbar    {\kern 0.18em\overline{\kern -0.18em B}{}\xspace}

\def\Dbar    {\kern 0.2em\overline{\kern -0.2em D}{}\xspace}
\def\Dz      {\ensuremath{D^0}\xspace}
\def\Dzbar      {\ensuremath{\Dbar^0}\xspace}

\def\Kbar  {\kern 0.2em\overline{\kern -0.2em K}{}\xspace}

\def\Kz    {\ensuremath{K^0}\xspace}
\def\Kzb   {\ensuremath{\Kbar^0}\xspace}
\def\KzKzb {\ensuremath{\Kz \kern -0.2em - \kern -0.2em \Kzb}\xspace}
\def\KS    {\ensuremath{K^0_{\scriptscriptstyle S}}\xspace} 
\def\KL    {\ensuremath{K^0_{\scriptscriptstyle L}}\xspace} 
\def\KSKL {\ensuremath{\KS \kern -0.2em - \kern -0.2em \KL}\xspace}
\def\Kp    {\ensuremath{K^+}\xspace}

\def\pip  {\ensuremath{\pi^+}\xspace}

\def\pipm  {\ensuremath{\pi^\pm}\xspace}

\newcommand{\gevc}{\ensuremath{{\mathrm{\,Ge\kern -0.1em V\!/}c}}\xspace}
\newcommand{\mevc}{\ensuremath{{\mathrm{\,Me\kern -0.1em V\!/}c}}\xspace}
\newcommand{\gevcc}{\ensuremath{{\mathrm{\,Ge\kern -0.1em V\!/}c^2}}\xspace}
\newcommand{\mevcc}{\ensuremath{{\mathrm{\,Me\kern -0.1em V\!/}c^2}}\xspace}

\newcommand{\Dtokspi}{\ensuremath{D^{+}\to\KS\pip}\xspace}
\newcommand{\Dtoksk}{\ensuremath{D^{+}\to\KS\Kp}\xspace}
\newcommand{\Dstoksk}{\ensuremath{D_s^{+}\to\KS\Kp}\xspace}
\newcommand{\Dstokspi}{\ensuremath{D_s^{+}\to\KS\pip}\xspace}

\newcommand{\Dps}{\ensuremath{D_{(s)}}}

\newcommand{\Dtokkpi}{\ensuremath{D^{+}\to K^+ K^- \pip}\xspace}

\def\Title#1{\begin{center} {\Large {\bf #1} } \end{center}}

\begin{document}

\Title{Mixing and \CP Violation in Charm Decays at \babar}

\bigskip
\bigskip

\begin{raggedright}  

{\it Riccardo Cenci\index{Cenci, R.} (on behalf of the \babar collaboration)\\
Department of Physics\\
University of Maryland\\
20742 College Park, Maryland, USA}
\end{raggedright}

\section{Introduction}

Mixing and \CP violation (\CPV) are two well-established phenomena
in the strange and bottom quark sectors~\cite{Beringer:2012zz}, but are less known in the charm sector.
Mixing in this sector has been measured by multiple experiments~\cite{Beringer:2012zz},
although at the time of the CKM2012 conference there was no single measurement that exceeds $5 \sigma$ significance. 
Recently the LHCb~\cite{Aaij:2011in} and CDF~\cite{Collaboration:2012qw} Collaborations 
have reported evidence of \CPV in the difference of the integrated \CP asymmetries 
in the $\Dz \to K^+ K^-$ and $\Dz \to \pi^+ \pi^-$ channels. 
This result was unexpected at the current experimental precision, 
and it may be a manifestation of New Physics (NP) effects, 
although a Standard Model (SM) explanation cannot be ruled out~\cite{Isidori:2011qw_plus}. 
These measurements have renewed the interest of the community in charm physics as a sector 
in which to search for NP manifestations.
Additional information and corroboration of these observations in other channels 
are necessary to clarify if NP effects can be detected in such phenomena.

In this report we present a measurement of mixing, and searches for direct and indirect 
\CP asymmetries in the charm sector using the full \babar data sample of $469 \invfb$.
Searches for direct \CP asymmetries have been performed 
in the \Dtoksk, \Dstoksk, \Dstokspi, and \Dtokkpi decay channels~\cite{chconj},
with both model-dependent and model-independent approaches being exploited for the last mode.
The mixing measurement and the search for indirect \CP asymmetry have been performed with the
$\Dz \to K^+ K^-$, $\Dz \to \pi^+ \pi^-$, and $\Dz \to K^\pm \pi^\mp$ decay modes.

\section{Searches for Direct \CP Violation}

Direct \CPV occurs when the total decay amplitude of a meson to a specific final state is different
from that of the corresponding anti-particle to the charge-conjugate final state.
For charged mesons, this is the only possible source of \CP asymmetry.
The total amplitude is different when there are competing contributions 
from two amplitudes of similar magnitude, but with different strong and weak phases.
In describing charm decay modes, we can distinguish between Cabibbo-favoured (CF), 
Cabibbo-suppressed (CS), and doubly Cabibbo-suppressed diagrams (DCS). 
According to the SM, when a CF and a DCS diagram contribute to the same decay, \CPV is negligible, 
while when we have two CS diagrams, we can observe \CPV asymmetries up to a level of 0.1\%.
When the final state includes a \KS, there is an additional contribution due to \CPV in $\Kz-\Kzb$ mixing
at a level of $(+(-)0.332\pm 0.006)\%$~\cite{Beringer:2012zz,Grossman:2011zk}, 
when a \Kz(\Kzb) is present in the dominant diagram.
The measured quantity is the time-integrated asymmetry defined as:
\begin{equation}
A_{\CP}=\frac{\Gamma(D^+_{(s)}\to f)-\Gamma(D_{(s)}^-\to \overline{f})}
{\Gamma(D^+_{(s)}\to f)+\Gamma(D^-_{(s)}\to \overline{f})}.
\end{equation}
Other than the \CP asymmetry, the measured value has two additional contributions:
one from asymmetry in $c\overline{c}$ production ($A_{\FB}$), 
and a second due to detector-induced charge asymmetry in charged-particle track reconstruction.
The first contribution is odd in the cosine of the center-of-mass polar angle, $\cos\theta^*$, 
and so can be removed by averaging the asymmetry in corresponding positive and negative intervals of $\cos\theta^*$.
The second contribution is removed by weighting each candidate according to its momentum 
by a coefficient determined by using a high-statistics control sample for which \CP asymmetry is absent.

\subsection{The \Dtoksk, \Dstoksk, and \Dstokspi modes}

For the \Dtoksk, \Dstoksk, and \Dstokspi modes, 
detector-induced charge asymmetries in reconstruction are reduced by means of
a data-driven method, which was used previously in an analysis of \Dtokspi~\cite{delAmoSanchez:2011zza}.
Signal candidates are weighted according to the momentum of the track not coming from the \KS decay.
The weights are produced using a high-statistics control sample which is free of any physics-induced asymmetries.

A simultaneous Maximum Likelihood (ML) fit to the $\Dps^+$ and $\Dps^-$ invariant mass distributions
is carried out to extract the signal yields and the respective asymmetries in ten equal intervals of $\cos\theta^*$.
The total yield for the signal is $(159.4 \pm 0.8) \times 10^3$  for \Dtoksk,
$(288.2 \pm 1.1) \times 10^3$ for \Dstoksk, and $(14.33 \pm 0.31) \times 10^3$ for \Dstokspi.
The asymmetries in each pair of symmetric $\cos\theta^*$ intervals are averaged to yield five $A_{\CP}$ values.
The weighted average of these $A_{\CP}$ values is found to be
$(0.155 \pm 0.360)\%$ for \Dtoksk,
$(0.00 \pm 0.23)\%$ for \Dstoksk,
and $(0.6 \pm 2.0)\%$ for \Dstokspi,
where the errors are statistical only.

The dominant systematic uncertainties are from the statistics of the control sample used to correct
for the charge asymmetry (\Dtoksk and \Dstoksk), and the binning in $\cos\theta^*$ (\Dstokspi).
Additional corrections are applied to account for bias from the fit, from the particle identification selectors,
and from $\KS-\KL$ interference, as detailed in Table~\ref{tab_final}.
\begin{table}[htb]
\caption{Summary table for the $A_{\CP}$ measurements.
Uncertainties, where reported, are first statistical, and second systematic (\babar Preliminary).} 
\vspace{-0.3cm}
\begin{center} \footnotesize
\begin{tabular}{|l|c|c|c|}
\hline
  & \Dtoksk & \Dstoksk & \Dstokspi \\
\hline \hline
$A_{\CP}$ value from the fit      & $(0.16 \pm 0.36)\%$ & $(0.00 \pm 0.23)\%$ & $(0.6 \pm 2.0)\%$ \\ 
\hline \hline
Bias Corrections & \multicolumn{3}{|c|}{}\\
\hline 
Toy MC experiments & $+0.013\%$ & $-0.01\%$ & $-$ \\ 
PID selectors & $-0.05\%$ & $-0.05\%$ & $-0.05\%$ \\
\KSKL interference 
& $+0.015\%$ & $+0.014\%$ & $-0.008\%$ \\
\hline \hline
$A_{\CP}$ corrected value & $(0.13 \pm 0.36 \pm 0.25)\%$ & $(-0.05 \pm 0.23 \pm 0.24)\%$ & $(0.6 \pm 2.0 \pm 0.3)\%$\\
\hline \hline
$A_{\CP}$ contribution & \multirow{2}{*}{$(-0.332 \pm 0.006)\%$} & \multirow{2}{*}{$(-0.332 \pm 0.006)\%$} & 
\multirow{2}{*}{$(0.332 \pm 0.006)\%$} \\
from \KzKzb mixing &&&\\
\hline \hline
$A_{\CP}$ value (charm only) & $(0.46 \pm 0.36 \pm 0.25)\%$ & $(0.28 \pm 0.23 \pm 0.24)\%$ & $(0.3 \pm 2.0 \pm 0.3)\%$\\
\hline
\end{tabular}
\normalsize \label{tab_final} \vspace{-0.7cm} \end{center} \end{table}
The measured final $A_{\CP}$ value for each mode is also reported in Table~\ref{tab_final},
and in the last row we also report the values after taking into account the expected contribution due to \KzKzb mixing.
These results are consistent with zero and with the SM prediction.

\subsection{The \Dtokkpi mode}

For the \Dtokkpi mode, different methods have been exploited to measure $A_{\CP}$, 
and to look for non-zero asymmetries in specific regions of the Dalitz plot (DP). 
Detector-induced charge asymmetry in track reconstruction is reduced by weighting the signal candidates
according to the $\cos\theta^*$ of the $D^+$ candidate and its position in the DP. 
Weights are extracted from a control sample of $\epem \to \tau^+\tau^-$ events.

First $A_{\CP}$ was measured by integrating over the whole DP,
using a procedure similar to that described above for the other modes. 
The total yield for the signal is $(228 \pm 0.8) \times 10^3$  \Dtokkpi decays.
The final value for $A_{\CP}$ is $(0.37 \pm 0.30 \pm 0.15)\%$, 
where the first error is statistical and the second is systematic.

Then, using a model-independent analysis, the signal yield was obtained by fitting the mass distribution
separately in four different regions of the DP, separating regions where the dominant contribution
is given by resonances or not. 
The signal yields, the efficiencies, and the final values for $A_{\CP}$ 
in each region are given in Table~\ref{tab_dalitz}.
\vspace{-0.6cm}
\begin{table}[htb]
\caption{Summary table for $A_{\CP}$ measurements.
Uncertainties, where reported, are first statistical, and second systematic
(\babar Preliminary).} 
\vspace{-0.3cm}
\begin{center}
\footnotesize
\begin{tabular}{|l|c|c|c|c|c|}
\hline
DP region & $N(D^+)$ & $\varepsilon (D^+)[\%]$ & $N(D^-)$ & $\varepsilon (D^-)[\%]$ & $A_{\CP}[\%]$ \\ 
\hline\hline
Below $\overline{K}^*(892)^0$ & $1882 \pm 70$ & $7.00$ & $1859\pm90$ & $6.97$ & $-0.7 \pm 1.6 \pm 1.7$\\
\hline
$\overline{K}^*(892)^0$ & $36770 \pm 251$ & $7.53$ & $36262 \pm 257$ & $7.53$ & $-0.3 \pm 0.4 \pm 0.2$ \\
\hline
$\phi(1020)$ & $48856 \pm 289$ & $8.57$ & $48009\pm289$ & $8.54$ & $-0.3 \pm 0.3 \pm 0.5$ \\
\hline
Above $\overline{K}^*(892)^0$ & 
\multirow{2}{*}{$25616 \pm 244$} & \multirow{2}{*}{$8.01$} & \multirow{2}{*}{$24560\pm242$} & 
\multirow{2}{*}{$8.00$} & \multirow{2}{*}{$1.1\pm 0.5\pm 0.3$} \\
and $\phi(1020)$ &&&&&\\ 
\hline 
\end{tabular}
\normalsize \label{tab_dalitz} \vspace{-0.5cm} \end{center} \end{table} 

Two alternative model-independent analyses using were performed.
The first used an adaptive binning of the DP, where each interval is equally populated, 
and the second used Legendre polynomial moments~\cite{Aubert:2008yd}. 
All the measured quantities sensitive to \CPV effects are consistent with zero.

Lastly, an additional analysis was performed which used an isobar model to describe the DP distribution 
by a coherent sum of amplitudes. Each resonance is parametrized with a different amplitude and phase
for $D^+$ and $D^-$, for a total of four parameters. A simultaneous fit to the DP distributions
was performed for $D^+$ and $D^-$ decays, and all measured \CPV parameter values are compatible with zero.

\section{Mixing and \CPV from a Lifetime Ratio Analysis}

The parameters that describe $\Dz - \Dzbar$ oscillations are proportional to 
the difference of masses ($m_i$) and widths ($\Gamma_i$) of the mass eigenstates:
$x \equiv (m_1 - m_2)/\Gamma$ and $y \equiv (\Gamma_1 - \Gamma_2)/2\Gamma$, where $\Gamma \equiv (\Gamma_1 + \Gamma_2)/2$.
Mixing will occur if the mass eigenstates differ from the flavour eigenstates, that is, if either $x$ or $y$ is non-zero. 
At present, experimental measurements are in agreement with SM predictions.
Unfortunately the corresponding theoretical predictions are affected by large computational uncertainties 
on the dominant long-range-diagram contributions, and this prevents these measurements from providing 
strong tests of the SM.
We perform a measurement of the mixing parameter $y_{\CP}$~\cite{Beringer:2012zz} 
and the \CP-violating parameter $\Delta Y$, defined as:
\begin{equation}
y_{\CP} \equiv  \frac{\tau_D}{2} \left ( \frac{1}{\tau^+} + \frac{1}{\overline{\tau}^+} \right) -1,
\quad \textrm {and} \quad
\Delta Y \equiv \frac{\tau_D}{2} \left ( \frac{1}{\tau^+} - \frac{1}{\overline{\tau}^+} \right),
\label{eq:ydelta}
\end{equation}
where ${\tau}^+$($\overline{\tau}^+$) is the effective lifetime of the \Dz(\Dzbar) 
when reconstructed in a CP-even eigenstate,
and ${\tau}_D$ is the average lifetime of the \Dz and \Dzbar 
when reconstructed in $\Dz \to K^\pm \pi^\mp$.
In the absence of \CPV, $y_{\CP} = y$ and $\Delta Y = 0$.

For this analysis five signal channels were used:
$\Dz \to K^+ K^-$, $\Dz \to \pi^+ \pi^-$, and $\Dz \to K^\pm \pi^\mp$ 
(flavour-tagged by reconstructing the $D^* \to \Dz \pipm$ decay);
$\Dz \to K^+ K^-$ and $\Dz \to K^\pm \pi^\mp$ (flavour-untagged).
A preliminary fit to the \Dz mass distributions was performed to extract yields and lifetime 
probability density functions (PDFs) for the signal and sideband regions, 
where only combinatorial background is present.
The same fit was performed to Monte-Carlo simulated events to extract yields and lifetime PDFs 
for the backgrounds resulting from misreconstructed charm decays.
The final simultaneous ML fit to the proper time and proper time error
distributions was then performed using only candidates from the signal region,
and this yielded ${\tau}^+ = (405.69 \pm 1.25)\fs$, $\overline{\tau}^+ = (406.40 \pm 1.25)\fs$, 
and ${\tau}_D = (408.97 \pm 0.24)\fs$.
Using Eq.~(\ref{eq:ydelta}), the values $y_{\CP} = (0.720 \pm 0.180 \pm 0.124)\%$ 
and $\Delta Y = (0.088 \pm 0.255 \pm 0.058)\%$ were obtained, 
where the first error is statistical and the second is systematic.

This measurement represents the most precise measurement of $y_{\CP}$, 
and excludes the no-mixing hypothesis at $3.3 \sigma$ significance.
The value of $y_{\CP}$ presented here is compatible with the previous \babar results~\cite{Aubert:2007en_plus}.
This measurement favors a lower value for $y_{\CP}$, 
and approaches the value of the mixing parameter $y$ when measured directly~\cite{Beringer:2012zz}, 
as expected if \CP is conserved. 
No evidence of \CPV was found.

\section{Conclusions and Acknowledgements}

In conclusion, an increase in precision and the inclusion of more analysis channels are needed 
in order to understand the origin of the \CPV in the charm sector
recently reported by LHCb~\cite{Aaij:2011in} and CDF~\cite{Collaboration:2012qw}.
We have searched for \CP-violating effects using the full \babar data sample,
and have reached a precision of $\mathcal{O}(10^{-3})$.
We have found no evidence of direct or indirect \CPV in any of the measured channels.
We have measured $y_{\CP}$ with the highest precision to date, 
and have excluded the no-mixing hypothesis at $3.3\sigma$ significance.

We are grateful for the excellent luminosity and machine conditions
provided by our \pep2\ colleagues, 
and for the substantial dedicated effort from
the computing organizations that support \babar.
The collaborating institutions wish to thank 
SLAC for its support and kind hospitality. 
This work is supported by
DOE
and NSF (USA),
NSERC (Canada),
CEA and
CNRS-IN2P3
(France),
BMBF and DFG
(Germany),
INFN (Italy),
FOM (The Netherlands),
NFR (Norway),
MES (Russia),
MICIIN (Spain),
STFC (United Kingdom). 
Individuals have received support from the
Marie Curie EIF (European Union)
and the A.~P.~Sloan Foundation (USA).


\def\Discussion{
\setlength{\parskip}{0.3cm}\setlength{\parindent}{0.0cm}
     \bigskip\bigskip      {\Large {\bf Discussion}} \bigskip}
\def\speaker#1{{\bf #1:}\ }
\def\endDiscussion{}
 
\end{document}

%% file: econfmacros.tex
\def\beq{\begin{equation}}
\def\eeq#1{\label{#1}\end{equation}}
\def\eeqn{\end{equation}}

\def\beqa{\begin{eqnarray}}
\def\eeqa#1{\label{#1}\end{eqnarray}}
\def\eeqan{\end{eqnarray}}

\let\bar=\overbar

\def\Dslash{\not{\hbox{\kern-4pt $D$}}}
\def\dslash{\not{\hbox{\kern-2pt $\del$}}}

\def\msb{{\bar{\ssstyle M \kern -1pt S}}}




%% file: article.bbl
\begin{thebibliography}{99}
\bibitem{chconj}
  The use of charge-conjugate decay modes is implied throughout this report.
\end{thebibliography}
